\documentclass[12pt]{iopart}
\usepackage{iopams}
\usepackage{graphicx}
\usepackage{dcolumn}
\usepackage{bm}
\usepackage{paralist}

\begin{document}

\title[Tipping Diffusivity in Information Accumulation Systems]{Tipping Diffusivity in Information Accumulation Systems: More Links, Less Consensus}

\author{J. K. Shin$^1$, and J. Lorenz$^2$}
\address{$^1$School of Mechanical Engineering, Yeungnam University, Kyongsan, South Korea}
\address{$^2$Chair of Systems Design, ETH Zurich, Kreuzplatz 5, 8032 Z\"urich, Switzerland}
\eads{\mailto{jkshin@yu.ac.kr}, \mailto{post@janlo.de}}



\begin{abstract}
Assume two different communities each of which maintain their respective opinions mainly because of the weak interaction between the two communities. In such a case, it is an interesting problem to find the necessary strength of inter-community interaction in order for the two communities to reach a consensus. In this paper, the Information Accumulation System (IAS) model is applied to investigate the problem. With the application of the IAS model, the opinion dynamics of the two-community problem is found to belong to a wider class of two-species problems appearing in population dynamics or in competition of two languages, for all of which the governing equations can be described in terms of coupled logistic maps. Tipping diffusivity is defined as the maximal inter-community interaction that the two communities can stay in different opinions. For a problem with simple community structure and homogeneous individuals, the tipping diffusivity is calculated theoretically. As a conclusion of the paper, the convergence of the two communities to the same value is less possible the more overall interaction, intra-community and inter-community, takes place. This implies, for example, that the increase in the interaction between individuals caused by the development of modern communication tools, such as facebook and twitter, does not necessarily improve the tendency towards global convergence between different communities. If the number of internal links increases by a factor, the number of inter-community links must be increased by an even higher factor, in order for consensus to be the only stable attractor.
\end{abstract}

\maketitle

\section{Introduction}\label{sec:intro}

Often social networks are composed of weakly connected clusters, or communities. The inter-community links between individuals are less than those within the community. When the inter-community interactions are few, the opinions can differ from one community to another. In such a case, it is an interesting question to ask under what condition the two communities reach a common consensus or can maintain different opinions. In \cite{Lambiotte.Ausloos.ea2007Majoritymodelnetwork}, this problem is treated using a local majority rule on randomly chosen triplets applied to a system composed of two fully connected core communities and some inter-community links. In the present paper, we apply the Information Accumulation System (IAS) model \cite{Shin2009Informationaccumulationsystem} to study essentially the same situation of two internally highly connected communities with a comparably low number of inter-community links. 

The IAS model describes a dynamic process through which information accumulates in an agent. Information can represent opinion, knowledge or also emotionality for a specific issue and is described as a real number from the interval [-1,1]. Thus, IAS is a kind of continuous opinion dynamics model \cite{Lorenz2007ContinuousOpinionDynamics,Hegselmann.Krause2002OpinionDynamicsand,Deffuant.Neau.ea2000MixingBeliefsamong}. In these models the evolution of opinions is invariant to additive shifts of the opinion space. If all initial opinions were increased by a constant, dynamics evolve essentially the same, just always shifted by that amount. This shift-invariance is not the case in the IAS model, where the extreme opinions -1 and +1 serve as natural extremes of opposite opinions.
Different signs of information represent two opposing opinions for a specific issue. Thus, in contrast to binary models of opinion formation  \cite{Lambiotte.Ausloos.ea2007Majoritymodelnetwork,Galam2008Sociophysicsreviewof,KatarzynaSznajd-Weron2000OpinionEvolutionin,Schweitzer.Holyst2000Modellingcollectiveopinion} in which just two opinion states can be represented, IAS can also represent the relative strength towards one of the two  opinions. For example +0.9 can represent ``strong preference for Coca-cola'' while -0.1 represent ``weak preference for Pepsi''. The state zero represents a natural ground state with no preference. In that sense our model is similar to the ``Continuous Opinions and Discrete Actions'' (CODA) model \cite{Martins2008ContinuousOpinionsand,Martins2008Mobilityandsocial}, where a continuous opinion is transformed to a discrete action by its sign. For a general overview on opinion dynamics models in physics see \cite{Castellano.Fortunato.ea2009Statisticalphysicsof}. Our two community case has also relations to studies of competition between two species in population dynamics \cite{Gilpin.Ayala1973Globalmodelsof} or two languages, as in \cite{Abrams.Strogatz2003Modellingdynamicsof,Pinasco.Romanelli2006CoexistenceofLanguages,Kosmidis.Halley.ea2005Languageevolutionand}. One underlying theoretical question is the understanding of coupled logistic maps \cite{Gilpin.Ayala1973Globalmodelsof,Satoh.Aihara1990Numericalstudycoupled-Logistic}. 

The dynamics of the information an agent holds in the model is dominated by two factors, the volatility of its old information, and information input of parts of the information of its neighbors. The \emph{volatility} of the old information quantifies the rate of forgetting of information. The strength of the impact of an agent on the opinions of others is called its \emph{diffusivity}.

The dynamical forces of the IAS model make it also a natural candidate for a simple model of collective emotions. In a dimensional approach on the representation of emotion usually the dimension of valence (``good'' versus ``bad'') turns out, after a factor analysis, to be most dominant (see \cite{Cowie2006EmotionallifeTerminological,Mauss.Robinson2009Measuresofemotion}). This can be naturally represented on the scale $[-1,1]$ with zero as the neutral ground state. Thus, the IAS model is of interest in the understanding of emotions as a collective social phenomenon \cite{Parkinson.Fischer.ea2005Emotioninsocial}. 

In the following we present a simplified version of the IAS model in Section \ref{sec:ias}, which we restrict to the two community case under small news. We consider that agents in a community have the same number of intra-community links towards members of the same community and the same lesser number of inter-community links towards members of the other community. This implies uniform intra-community diffusivity and uniform inter-community diffusivity of information. The dynamical behavior in this case is fully characterized analytically. The system has different attractors depending on the volatility, the intra- and the inter-community diffusivity. For high volatility naturally only the ground state zero is attractive. With rising diffusivities two other attractors evolve, one, where both communities agree on a positive opinion and an analog one where both agree on a negative opinion. There also exists a region where attractors of opposing opinions of the two communities exist. Section \ref{sec:lang} applies the IAS model to the question of two coexisting languages. The most interesting fact is that the coexistence of opposing opinions or languages can become attractive and stable when the overall diffusivity rises even when the ratio of inter- versus intra-community diffusivity stays constant. This implies that modern diffusivity-enhancing communication technologies need not raise the tendency towards global convergence in opinion or language between different communities. On the contrary, the ratio of the number of inter- versus intra-community links must be higher to ensure convergence to a consensual state.

\section{Information Accumulation System Model with two Communities}\label{sec:ias}

A simplified version of the IAS is expressed in terms of an iterative map as follows. Consider $n$ agents. The information of agent $i$ at time step $t$ is labeled $y_{i}^t$. With time running in discrete steps agent $i$ updates her information by the update map
\begin{equation}
 y_{i}^{t+1}=(1-\Delta) y_{i}^t + \sum_{j\in \Gamma_i}\omega y_{j}^t (1-|y_{i}^t|) \label{eq:1}
\end{equation}
Eq.~(\ref{eq:1}) shows that the information of agent $i$ at time step $t+1$ is made up from two different contributions. The inheritance term $(1-\Delta)y_{i}^t$  where $0< \Delta \leq 1$ is the \emph{volatility}. When the rest of the terms are neglected, information will decay to the ground state zero. Thus, $\Delta$ can be seen as a rate of forgetting the old information (``Volatility'' is meant here in the thermodynamic sense of information ``vaporizing'' to the ground state, and not in the stochastic sense quantifying the standard deviation of noise as used in finance.) 

The second contribution to the information at the next time step is the diffusivity term $\sum_{j\in \Gamma_i} \omega y_{i}^t$, which reflects the interactions between the agents in the system. The set $\Gamma_i$ is the set of neighbors of agent $i$. The parameter $0\leq\omega<1$ is called \emph{diffusivity}. Upon interaction, a neighbor $j$ delivers part of its information, a fraction of $\omega$ to agent $i$. Thus, it can be seen as the loudness of the voices of the neighbors. The factor $(1-|y_{i}^t|)$ is a saturation factor. It guarantees that the size of the information is limited to ${|y_{i}^t|\leq 1}$ at all time steps (when some bounds on the parameter space and the initial conditions are respected). 

The simplified version is derived from a general form of the IAS model presented in \cite{Shin2009Informationaccumulationsystem}. The derivation is presented in \ref{ap:2} together with some more details about the interpretation of terms in different contexts. In the following we restrict the model to a two community situation and fully characterize its dynamics with respect to initial conditions, volatility, inter-, and intra-community diffusivity. 

Assume a social network with two communities and every individual in the system having $m_0$ intra-community and  $m_X$ inter-community neighbors. Thus, weak inter-community interaction is described by $m_0>m_X$. We further assume that initial information of all the individuals in the same community is the same. With this assumption, $y_{1}^t,\;y_{2}^t$ represent the information level of the individuals in the communities 1 and 2 at time step $t$. 

With the definition of \emph{inter-community diffusivity} and \emph{intra-community diffusivity} as
\begin{eqnarray}
 \Omega_0=m_0\omega,\quad \Omega_X=m_X\omega, \nonumber
\end{eqnarray}
Eq.~(\ref{eq:1}) simplifies to the following system
\begin{eqnarray}
y_{1}^{t+1}&=\tau y_{1}^t+(\Omega_0 y_{1}^t+\Omega_X y_{2}^t)(1-|y_{1}^t|) \nonumber \\
y_{2}^{t+1}&=\tau y_{2}^t+(\Omega_X y_{1}^t+\Omega_0 y_{2}^t)(1-|y_{2}^t|). \label{eq:2}
\end{eqnarray}

\begin{figure*}
 \includegraphics[width=\textwidth]{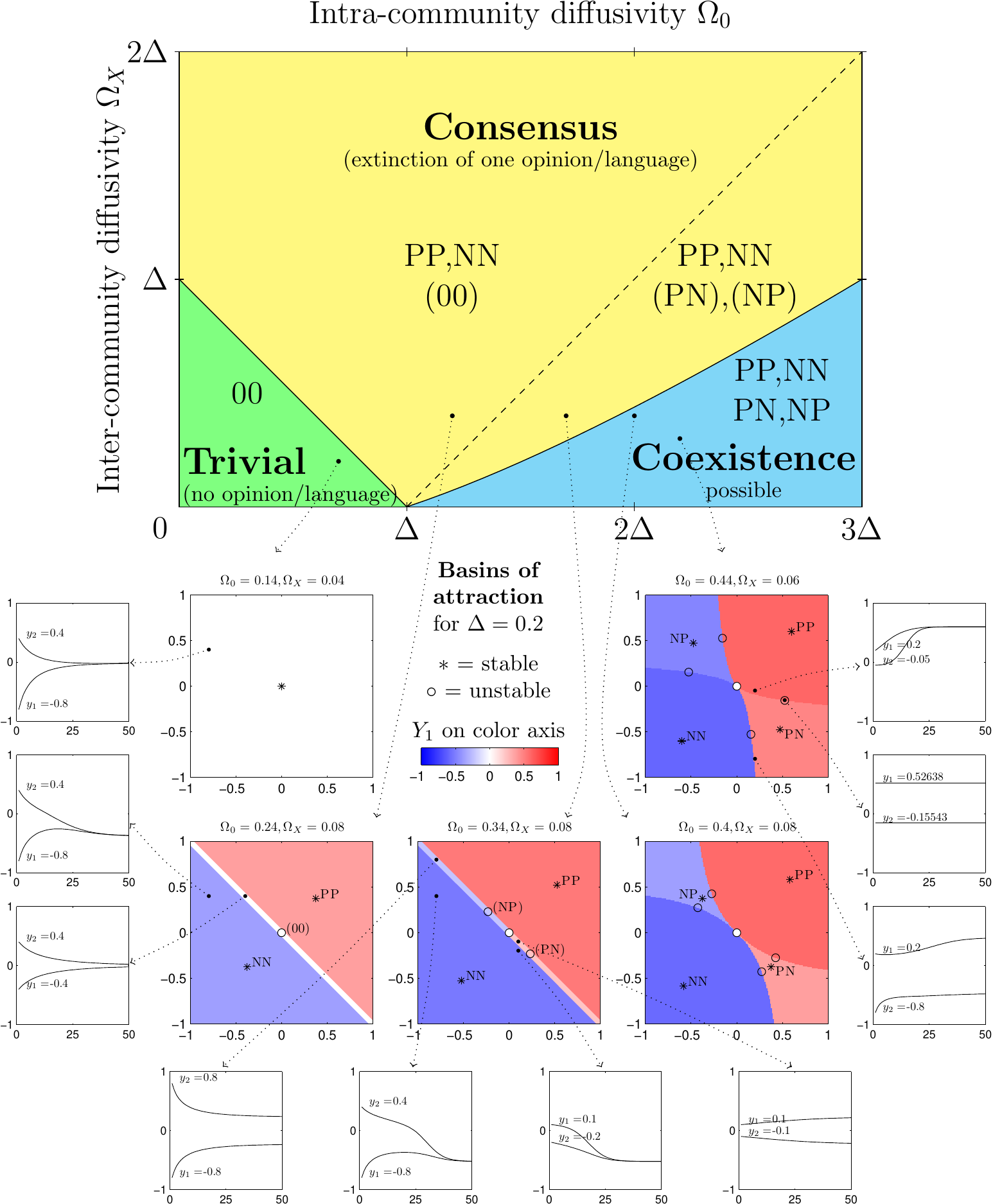}
\caption{\textbf{(top)} Phase diagram in the \mbox{$(\Omega_0,\Omega_X)$-plane}, parameterized by $\Delta$. Note, that the system cannot be regarded as having community structure in the region $\Omega_X>\Omega_0$. The region is included for theoretical reasons. \textbf{(center)} Basins of attractions for all attractive fixed points in the $(y_1,y_2)$-plane for certain $(\Omega_0,\Omega_X)$ and $\Delta=0.2$. The color-axis shows the value of $Y_1$ of the corresponding fixed point. The corresponding value of $Y_2$ is either equal to $Y_1$ in PP/NN-mode, or equal to $-Y_1$ in PN/NP-mode. There are two basins of attraction for different values in the coexistence zone to demonstrate how borders change.  \textbf{(periphery)} Exemplary trajectories for initial conditions $(y_{1}^0,y_{2}^0)$ as marked in the basins of attractions.}
\label{fig:2}
\end{figure*}

For convenience, let us define the information modes, of the system depending on the signs of $y_{1,i}$ and $y_{2,i}$. For example, PN-mode denotes positive $y_{1}^t$ and negative $y_{2}^t$. Analog, PP-, NN-, and NP-mode for different signs in of $y_{1}^t$ and $y_{2}^t$ being both positive, both negative, or negative and positive. For a system in PN-mode at time step $t$, Eq.~(\ref{eq:2}) can be written as
\begin{eqnarray}
y_{1}^{t+1}&=\tau y_{1}^t+(\Omega_0 y_{1}^t+\Omega_X y_{2}^t)(1-y_{1}^t) \nonumber \\
y_{2}^{t+1}&=\tau y_{2}^t+(\Omega_X y_{1}^t+\Omega_0 y_{2}^t)(1+y_{2}^t). \label{eq:3}
\end{eqnarray}
Similarly, for the PP-mode, the following equations hold. 
\begin{eqnarray}
y_{1}^{t+1}&=\tau y_{1}^t+(\Omega_0 y_{1}^t+\Omega_X y_{2}^t)(1-y_{1}^t) \nonumber \\
y_{2}^{t+1}&=\tau y_{2}^t+(\Omega_X y_{1}^t+\Omega_0 y_{2}^t)(1-y_{2}^t). \label{eq:4}
\end{eqnarray} 
The equations for NP-mode and NN-mode are analog with switched signs in the last factor of the last summand in both equations. Given initial conditions $y_{1}^0,\ y_{2}^0$, the iterative maps given in Eqs.~(\ref{eq:3}) and (\ref{eq:4}) can be updated for later time steps. One of the key interests of the present study is on the information modes of the steady state solutions. The evolving mode depends on the three key parameters of the system, $\Delta$, $\Omega_{0}$, and $\Omega_{X}$. The steady-state solution depends also on the initial condition. For example, an initial condition $y_{1}^0>0,\ y_{2}^0>0$ always drives the system to a PP-mode while $y_{1}^0>0,\ y_{2}^0<0$ may drive it to a PN-mode (but it need not). Fig.~\ref{fig:2} shows ten different trajectories embedded in a general picture of the dynamical behavior.  

The steady state solutions can be found by iteration of updating maps, with random initial conditions. Another convenient method is solving the fixed point equations obtained by setting  $y_{1}^{t+1}=y_{1}^t = Y_1$ and  $y_{2}^{t+1}=y_{2}^t= Y_2$ in Eqs.~(\ref{eq:3}) and (\ref{eq:4}). Then,  $(Y_1,Y_2)$ denotes a fixed point. The equations for the fixed points for the PN-mode are obtained from the maps for the PN-mode given in Eq.~(\ref{eq:3}),
\begin{eqnarray}
Y_1&=(1-\Delta)Y_1+(\Omega_0Y_1+\Omega_XY_2)(1-Y_1) \nonumber \\
Y_2&=(1-\Delta)Y_2+(\Omega_XY_1+\Omega_0Y_2)(1+Y_2). \label{eq:5}
\end{eqnarray}
It can be easily shown that Eq.~(\ref{eq:5}) has a trivial solution $Y_1=Y_2=0$ and a skew-symmetric solution of $Y_1=-Y_2$. The skew symmetric solution defined as $Y_\mathrm{PN} = Y_1=-Y_2>0$ is easily obtained from Eq.~(\ref{eq:5}):
\begin{equation}
Y_\mathrm{PN}=1-\frac{\Delta}{\Omega_0-\Omega_X}, \quad \mathrm{when }\ \Omega_0-\Omega_X \geq \Delta. \label{eq:6}
\end{equation} 
Notice, that under certain conditions also another pair of fixed points exists in PN-mode. They have lengthy analytical form and are unstable and not attractive, therefore they are reported in \ref{ap:1}.

\begin{figure}\centering\centering
 \includegraphics[width=0.8\columnwidth]{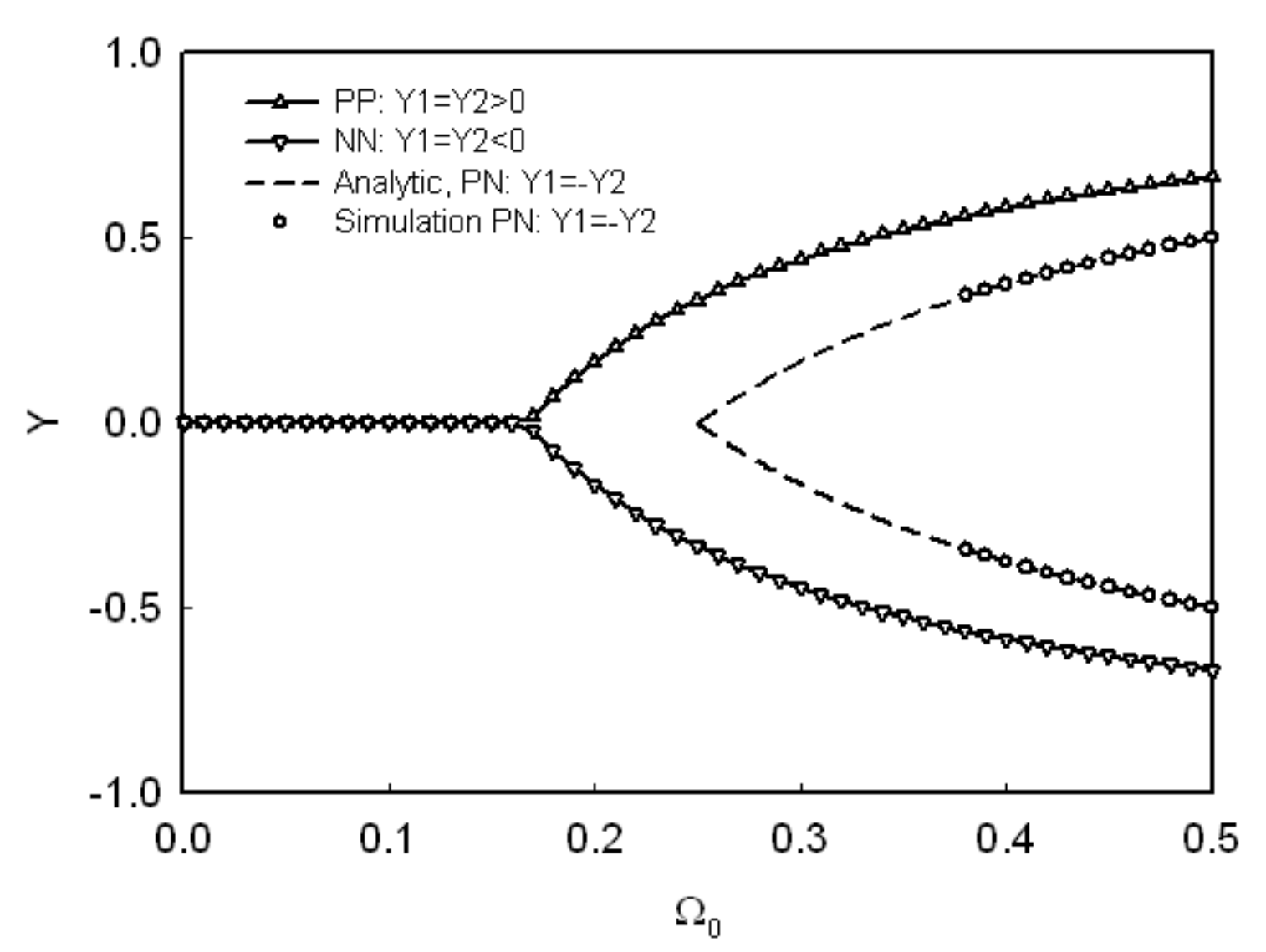}
\caption{Solutions obtained for the case  $\Delta =0.2,\;\Omega_X=0.2\Omega_0$. Lines denote analytic solutions and symbols denote stable solutions which can be obtained by numerical computation with random initial values. }\label{fig:1}
\end{figure}

Following the same process, we can find the analytic solutions for the PP-mode. The PP-mode also has the trivial solution and a unanimous solution $Y_1=Y_2 = Y_\mathrm{PP}$. The unanimous solution is 
\begin{equation}
Y_\mathrm{PP}=1-\frac{\Delta}{\Omega_0+\Omega_X}, \quad \mathrm{when }\ \Omega_0+\Omega_X\geq\Delta. \label{eq:7}
\end{equation} 
If the conditions (\ref{eq:6}) and (\ref{eq:7}) are not satisfied, only the trivial solution is possible. No other solutions but the explained were found from the PN-mode and PP-mode equations. Table \ref{tab:App} in the Appendix gives a concise overview of all these fixed points. 

Fig.~\ref{fig:1} shows all the steady state solutions that can be obtained for the case $\Delta =0.2,\;\Omega_X=0.2\Omega_0$. (We omit only the unstable fixed points mentioned in \ref{ap:1}.) The condition $\Omega_X=0.2\Omega_0$ is taken as an example case. For the small diffusivity case of $\Omega_0+\Omega_X<\Delta $, or  $\Omega_0<0.167$, only the trivial solution is possible. For $\Omega_{{0}}\ge 0.167$, both the PP-mode and NN-mode are possible. In PP-mode the solution becomes $Y_1=Y_2=Y_\mathrm{PP}$ and in NN-mode $Y_1=Y_2=-Y_\mathrm{PP}$.
PN-mode is possible at a higher intra-community diffusivity and small enough inter-community diffusivity. The condition for the PN-mode is  $\Omega_0-\Omega_X\ge \Delta$ or  ${\Omega_0\ge 0.25}$. It is important to observe that there exists a parameter range in which solutions in PN-mode are unstable but attractive on a thin basin of attraction. Thus, they can only be observed analytically and not in simulations. 

\begin{figure}\centering
 \includegraphics[width=0.8\columnwidth]{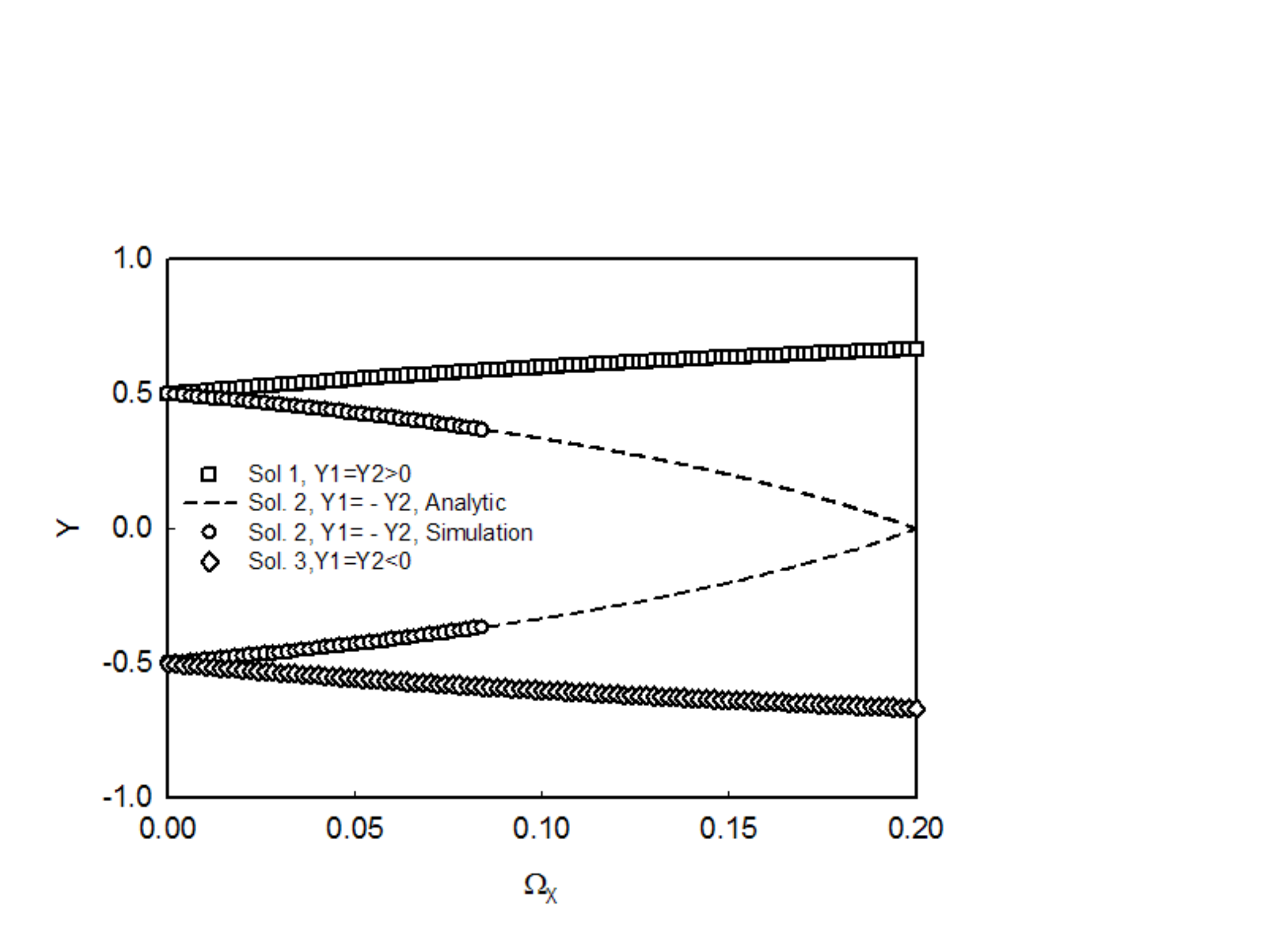}
\caption{Dependence on inter-community diffusivity ($\Delta=0.2$, $\Omega_0=2.5\Omega_X \Leftrightarrow \Omega_X=0.4\Omega_0$). For a given intra-community diffusivity, coexistence (PN- or NP-mode) is possible only when the inter-community diffusivity is below some threshold. }\label{fig:3}
\end{figure}

Solutions in PN-mode can also be stable. For example, Fig.~\ref{fig:3} shows stable solutions in PN-mode for $\Omega_X\le 0.088$ for the case of $\Delta=0.2$ and $\Omega_0=0.4$. Let us consider a critical inter-community diffusivity  $\Omega_X^T=\Omega_X^T(\Delta ,\Omega_0)$ defined as the maximum value of $\Omega_X$ under which solutions in PN-mode can be stable. It can be obtained analytically through an eigenvalue analysis for the linearized system of Eq.~(\ref{eq:3}). The critical inter-community diffusivity is obtained for the condition that the eigenvalue equals one. The result is 
\begin{equation}
\Omega_X^T=\Omega_0+\frac{1}{2}\Delta-\frac{1}{2}\sqrt{\Delta^2+8\Omega_0\Delta }.
\label{eq:OmegaXC}
\end{equation}

Fig.~\ref{fig:2} shows a phase diagram of solutions in \mbox{$(\Omega_0,\Omega_X)$-plane} with axis' parametrized by $\Delta$. The parameterization of the $(\Omega_0,\Omega_X)$-plane in units of $\Delta$ is general because for joint scaling e.g. by a positive constant $c$ as $\Omega'_0,\Omega'_X,\Delta' = c\Omega_0,c\Omega_X,c\Delta$ all solutions and conditions are exactly the same because $c$ cancels out as it can be easily checked. The entire parametric space can be divided into two regions, a trivial one and a non-trivial one, divided by the line $\Omega_0+\Omega_X=\Delta$ from Eq.~(\ref{eq:7}). The trivial region appears when the diffusivities are so small that no meaningful level of information can be accumulated. The only solution possible in this region is $(0,0)$. 

The non-trivial region can be further divided into two zones, the consensus zone and the coexistence zone. The largest part of the parametric space is occupied by the consensus zone in which only solutions in consensual modes (PP and NN) are stable. The trivial solution $(0,0)$ becomes unstable, though it is still attractive on the anti-diagonal of the $(y_1,y_2)$-plane ($y_1=-y_2$) when $\Omega_0-\Omega_X<\Delta$. For $\Omega_0-\Omega_X>\Delta$ the trivial solution $(0,0)$ looses attractivity and two unstable solutions in PN/NP-mode evolve which are again attractive each on its half of the anti-diagonal of the \mbox{$(y_1,y_2)$-plane}. Unstable solutions are neglected in classifying the zones, because they are only of theoretical interest. The line $(\Omega_0,\Omega_X^T(\Omega_0))$ divides the non-trivial region into the consensus zone  and the coexistence zone. The coexistence zone appears when the inter-community diffusivity is relatively low. In the coexistence zone, both the consensual (PP and NN) and the coexistence (NP and PN) modes are possible. The realization of each depends on initial conditions. 

Fig.~\ref{fig:2} also shows five basins of attractions of fixed points in the $(y_1,y_2)$-plane for certain values in the $(\Omega_0,\Omega_X)$-plane when $\Delta=0.2$. The basins of attraction were computed numerically for a fine grid in the $(y_1,y_2)$-plane. The structure of the zones appear to be the same for different values of $\Delta$.  Further on, some trajectories are shown and attached to certain points in the basins of attraction, which mark their initial conditions.

In order for the two communities to maintain different opinions, the inter-community diffusivity should be below $\Omega_X^T$. Consider two mutually isolated communities that are inheriting their own opinion or social norm for years. Now what will happen if the two communities begin to interact with each other? When the interaction is still low enough, they can maintain their respective style of life. When the inter-community interaction is increased above the critical level, a small perturbation of the PN-equilibrium can lead to convergence to a PP/NN-equilibrium. Thus, one of the two communities may discard their traditional opinion to adopt the foreign one. Tipping has occurred. For this reason, we call the critical inter-community diffusivity \emph{tipping diffusivity}. 

\begin{figure}\centering
\includegraphics[width=0.8\columnwidth]{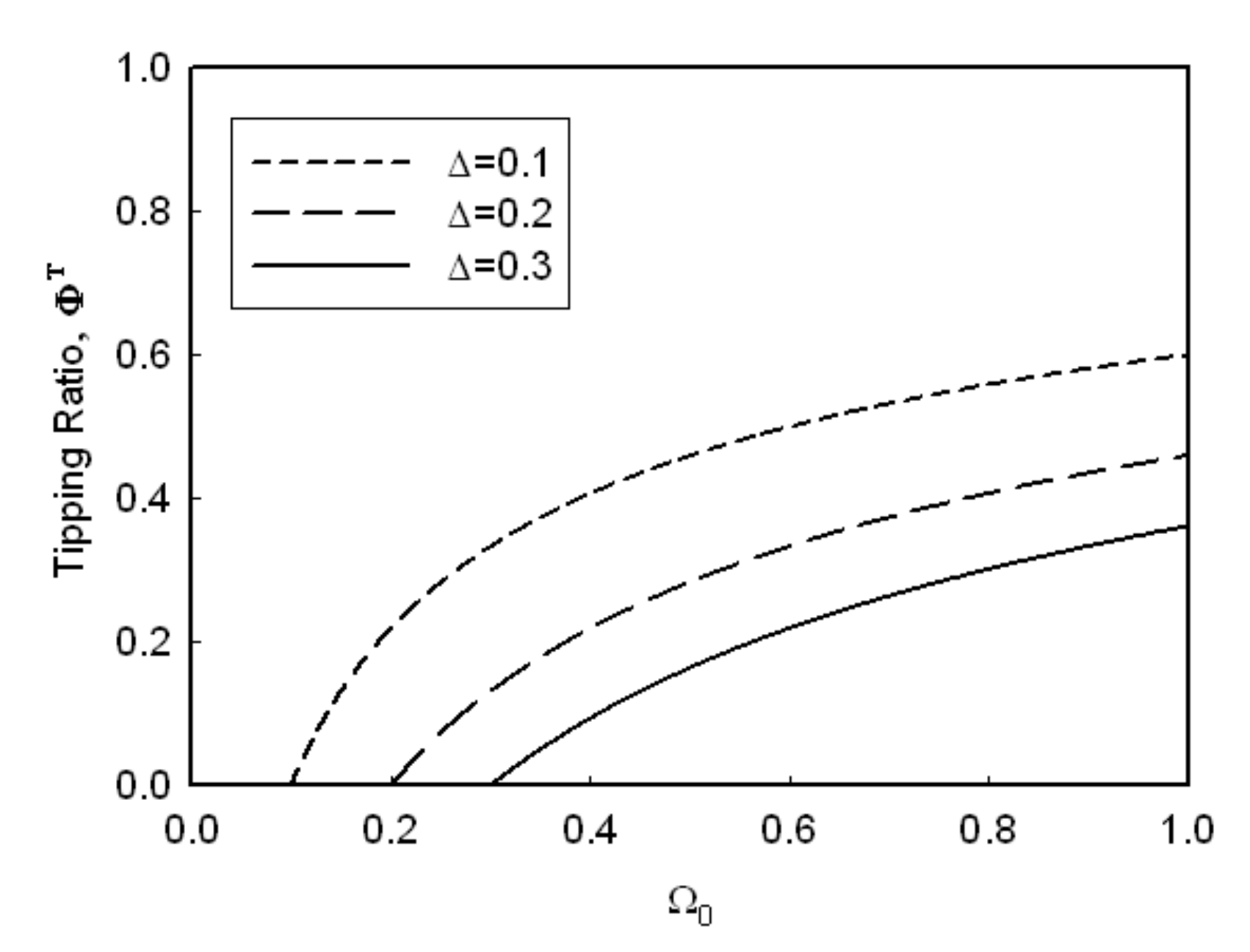}
\caption{Tipping ratio $\Phi^T$as it dependes on intra-community diffusivity $\Omega_0$.}\label{fig:4}
\end{figure}

In the following we study the consequences of a general increase in diffusivity, as we experience it through modern technologies such as internet and mobile phones. To that end, let us consider the diffusivity ratio $\phi=\frac{\Omega_X}{\Omega_0}$ which represents the ratio between the sizes of the inter-community diffusivity to that of the intra-community diffusivity. Naturally, it should hold $\phi<1$. Consequently, we define the tipping diffusivity ratio as
\begin{equation}
\Phi^T=\frac{\Omega_X^T}{\Omega_0},
\end{equation}
and study, how it depends on the inter-community diffusivity. Below the tipping diffusivity ratio, PN-mode is stable and thus coexistence of opinions is possible. In Fig.~\ref{fig:4}, $\Phi^T$ is plotted against the intra-community diffusivity $\Omega_0$ for three values of the volatility $\Delta$. In \cite{Lambiotte.Ausloos.ea2007Majoritymodelnetwork}, the tipping ratio is calculated as a ratio between the number of interface nodes to that of the core community nodes. In an example calculation, the tipping ratio was obtained at about 0.32, which does not depend on the characteristic of the individual agents. The value of 0.32 is quoted to compare with the results shown in Fig. \ref{fig:4}. The tipping ratio of the present study depends on the volatility and the inter-community diffusivity. We can observe: (i) The lower the volatility $\Delta$ the higher the tipping diffusivity ratio $\Phi^T$. (ii) The larger the intra-community diffusivity $\Omega_0$, the larger the tipping ratio $\Phi^T$.

Observation (i) is quite intuitive meaning that a society does not tip easily when information does not vanish easily. Observation (ii) means that an increase in the number of internal links, makes the society more resistant to tipping, even when the inter-community links are increased with the same ratio. That is what is meant with ``more links, less consensus'' in the title of this paper. Similar the tipping ratio rises when the individual diffusivity ($\omega$) increases while the network remains constant. In other words, tipping towards consensus gets less likely when voice get louder. 

It is evident that this result is a direct consequence of the two important characteristics of the IAS model. The first one is that the IAS is a continuous model. And the second one is that the diffusion term is an increasing function of the diffusivity. The model in \cite{Lambiotte.Ausloos.ea2007Majoritymodelnetwork} is not a continuous one, paying no attention to the dependence of the tipping ratio on the diffusivity or other factors. And no known model seems to have a similar diffusion term as in IAS, making the result unique.

The ``more links, less consensus''-effect in this study is interesting looking at globalization. Modern communication technologies increase the level of diffusivity, but this does not necessarily increase the tendency towards global convergence between different communities or civilizations. On the contrary, the result suggests that differences between civilizations could be strengthened, because the tipping diffusivity is higher even when intra- and inter-community diffusivity are increased with a constant ratio. 

It is interesting to note that the two community model formulated in terms of the IAS model is comparable to the two species model appearing in population dynamics, in which the governing equation is a set of coupled logistic maps \cite{Gilpin.Ayala1973Globalmodelsof}. One can show that Eq.~(\ref{eq:5}) in the present study can be transformed into a set of symmetric coupled logistic maps. The behavior of the coupled logistic maps strongly depends on the type of the coupling \cite{Gilpin.Ayala1973Globalmodelsof,Satoh.Aihara1990Numericalstudycoupled-Logistic}. The type of the coupling for the present study is not a simple one so that all the analysis in this paper were completed without utilizing the knowledge of the logistic map. Anyhow, the IAS model made a connection of opinion dynamics to well known problems in population dynamics. For example, we could also apply the IAS model for a two-language competition problem \cite{Abrams.Strogatz2003Modellingdynamicsof,Pinasco.Romanelli2006CoexistenceofLanguages,Kosmidis.Halley.ea2005Languageevolutionand} for which a coupled logistic map can be applied \cite{Pinasco.Romanelli2006CoexistenceofLanguages}. This example follows in the next section.

\section{Application to Language competition} \label{sec:lang}

As a possible application of the two-community model described in this study, consider a language competition problem. After the seminal work of Abrams and Strogatz \cite{Abrams.Strogatz2003Modellingdynamicsof}, a number of papers were published on the language competition problem. The main purpose of these works is to predict the fate of two or more competing languages in a system. While the result from Abrams suggests that the two languages cannot coexist stably \cite{Abrams.Strogatz2003Modellingdynamicsof,Stauffer.Castello.ea2007MicroscopicAbrams-Strogatzmodel}, other studies predict that both of the languages can survive \cite{Pinasco.Romanelli2006CoexistenceofLanguages,Castello.Toivonen.ea2008Modellinglanguagecompetition,Patriarca.Leppaenen2004Modelinglanguagecompetition}, depending on conditions. For the latter cases, the mode of coexistence is either through bilingualism \cite{Castello.Toivonen.ea2008Modellinglanguagecompetition} or two monolinguals \cite{Pinasco.Romanelli2006CoexistenceofLanguages,Patriarca.Leppaenen2004Modelinglanguagecompetition}.

The language models can be classified in two types depending on the variables used to represent the dominance of the languages. In most of the models, as in the original model of Abrams and Strogatz, the governing equation is described in terms of the population size \cite{Pinasco.Romanelli2006CoexistenceofLanguages} or concentration of population  \cite{Abrams.Strogatz2003Modellingdynamicsof,Stauffer.Castello.ea2007MicroscopicAbrams-Strogatzmodel,Patriarca.Leppaenen2004Modelinglanguagecompetition}. The governing equation is formulated either in a mean field approximation or by agent-based models. These models are discrete ones in the sense that each of the individuals is classified as speaker of one of the two languages in competition, A or B. Bilingualism cannot be described in two status models. This is comparable to another type of model in which the language dominance is represented by the number of words the agents use \cite{Kosmidis.Halley.ea2005Languageevolutionand}. For example, an agent can have up 20 words, 10 words from each of the two languages A and B. If two agents with different vocabularies meet, each of them can update their vocabulary through a learn-and-forget process. An agent is not described simply as, a speaker of A or B.  Bilingualism can be represented. Actually, agent-based simulations resulted in most cases in a homogeneous bilingual society \cite{Kosmidis.Halley.ea2005Languageevolutionand}.

Similar to the work of \cite{Kosmidis.Halley.ea2005Languageevolutionand}, we can apply IAS to the language competition problem. We assume two communities 1 and 2, initially using language A and language B respectively. We assume that the total number of words in each of the languages is $V_w$. In general, an agent can have mixed vocabulary of words from A and B. We define the information to be used in IAS as 
\begin{eqnarray}
 y_1 & = (v_1^A-v_1^B)/V_w \nonumber \\
 y_2 & = (v_2^A-v_2^B)/V_w. \label{eq:lang1}
\end{eqnarray}
In this equation, subscripts 1 and 2 denote the community, superscripts A and B denote the language, and $v$ denotes size of vocabulary. Thus, $v_1^B$ represents the number of words from B in the vocabulary of the agents in community 1. The size of vocabulary of the agents 1 and 2 is denoted as
\begin{eqnarray}
 V_1 = v_1^A + v_1^B \nonumber \\
 V_2 = v_2^A + v_2^B \label{eq:lang2}
\end{eqnarray}
The definition of $y_1$ and $y_2$ simply means the \emph{fraction of the excessive number of A-words over B-words} in the vocabulary of each of the agents 1 and 2, respectively. A positive (negative) value of $y_1$ and $y_2$ means that there are more (less) A-words in each of the agents. And it naturally follows that $|y_1|,|y_2|\leq 1$. An agent whose information level is very close to 0 is a bilingual person, or sometimes an agent without language. We can call a person with positive (negative) information as a person of language A (B), regardless of its community. 

Now let us apply the IAS model. One time step means one generation. And $y_{1}^t$ describes language usage of an average adult of society 1 at generation $t$. The volatility reflects the limited inheritence from the agent's parents. Diffusivity means the cumulative influence from the neighbors the agent interacts with through its lifetime. If an agent interacts with its parent or its neighbors with A-excessive vocabulary, it learns some part of the excessive vocabulary and its vocabulary will move towards A-excessive direction. Assume that the two societies 1 and 2 are initially separated from each other, which means that $\Omega_X=0$. Without inter-community diffusivity, the agents in community 1 (2) have vocabularies purely composed of A-words (B-words). The values of these two variables can be calculated easily and will be shown below for an example case. When the two communities begin interaction, the values of $y_1$ and $y_2$ will change, depending on the diffusivities $\Omega_0,\Omega_X$ as well as on the volatility $\Delta$. For relatively small values of $\Omega_X$, the two communities will reach the PN mode, which is a coexistence mode. But for larger values of $\Omega_X\geq\Omega_X^T$, tipping will occur and one of the two languages will go extinct. 

As an example case, assume the total number of words in language A and language B is given by $V_w = 100,000$, respectively. And $\Delta=0.2$, $\Omega_0+\Omega_X=0.3$,  for both of the societies and initially the members of community 1 only speak words of A (which implies $y_{1}^0>0$), while members of community 2 speak the same amount of words but only from the language B (which implies $y_{2}^0 = -y_{1}^0<0$).  We want to see the change with the increase in the inter-community diffusivity. Increase in inter-community diffusivity means decrease in intra-community diffusivity as we are assuming $\Omega_0+\Omega_X$ remains constantly $0.3$. 
 
For isolated societies with $\Omega_X=0$ (and thus $\Omega_0=0.3$), the system remains in PN-mode when it starts in PN-mode. The values of $y_1$ and $y_2$ are given by $y_1=-y_2=1-\Delta/\Omega_0 = 1 - 0.2/0.3 = 1/3$ (see Eqs.~(\ref{eq:6}) or (\ref{eq:7})). For the completely isolated case $v_1^B = v_2^A=0$ it follows from Eq.~(\ref{eq:lang1}) that an agent of community 1 has a vocabulary of size about $33,000$ A-words. Similarly agents of community 2 have vocabularies of about $33,000$ B-words. If volatility decreases or diffusivity increases, the vocabulary of the agents increases, too.

When the inter-community diffusivity is not zero but relatively small such that $0<\Omega_X<\Omega_X^T$ is satisfied, the solution is still in a PN-mode. Assume $\Omega_X=0.02$ (meaning $\Omega_0=0.28$). When $\Omega_0$ is given, we can calculate the tipping diffusivity from Eq.~(\ref{eq:OmegaXC}). For $\Omega_0=0.28$, it is $\Omega_X^T=0.031$. Also the solution can be easily obtained from Eq.~(\ref{eq:6}) as $y_1=-y_2=1-\frac{\Delta}{\Omega_0-\Omega_X}=1-\frac{0.2}{0.28-0.02}\approx0.23$. Now, the reduced value of $y_1$ compared to the isolated case means that the vocabulary of agents of community 1 is an unequal mix of words from the two languages. Eq.~(\ref{eq:lang1}) gives the following relation
\begin{eqnarray}
 v_1^A-v_1^B \approx 0.23 \times 100,000 = 23,000. \label{eq:lang3}
\end{eqnarray}
Agent 1 has in its vocabulary $23,000$ A-words more than B-words. To find the respective number of A-words and B-words in the vocabulary, another equation is needed. For this we assume that the size of vocabulary is maintained fixed as long as $\Omega_0+\Omega_X$ is constant. In this case, we have
\begin{eqnarray}
 V_1 = v_1^A+v_1^B \approx 33,000 \label{eq:lang4}
\end{eqnarray}
Solving Eq.~(\ref{eq:lang3}) and (\ref{eq:lang4})  together, we get $v_1^A \approx 28,000$ and $v_1^B\approx5,000$. Analog for community 2. 

Compared to the isolated case, each of the agents has replaced about $5,000$ of his domestic words with foreign words.
At a value of $\Omega_X=0.05$ (and $\Omega_0=0.25$), the tipping diffusivity is $\Omega_X^T = 0.018$. Tipping occurred and the solution is either in PP-mode or in NN-mode. At PP-mode, the solution is given from Eq.~(\ref{eq:7}), $y_1 = y_2 = 1 - \frac{\Delta}{\Omega_0+\Omega_X} = 1/3$, meaning about $33,000$ A-words only for all the agents in the system. 

Notice that we could apply the same procedure for the case $\Omega_0+\Omega_x\leq \Delta$. In this case we reach a vocabulary of size zero, which means no language. Thus, the trivial region corresponds to a primitive society in which there is essentially no language. It is remarkable that the solution $(0,0)$ is not a stable solution in the non-trivial region (see Fig.~\ref{fig:2}). This means that perfect bilingualism is never obtained in communities with language of vocabulary size larger than zero.  

In studies using Abrams and Strogatz type of models, the results show that the two competing languages cannot coexist \cite{Abrams.Strogatz2003Modellingdynamicsof,Stauffer.Castello.ea2007MicroscopicAbrams-Strogatzmodel} in general. But in some special cases, the coexistence was predicted to be possible. Abrams speculated that the language decline can be slowed down by strategies such as policy making, education and advertising to increase the prestige of the endangered languages \cite{Abrams.Strogatz2003Modellingdynamicsof}. In a spatial model, coexistence mode was obtained only within a narrow zone around the geometrical border of the two interacting societies \cite{Patriarca.Leppaenen2004Modelinglanguagecompetition}. In terms of IAS, Abrams strategy, especially the education, is equivalent to increasing the intra-community diffusivity, thus increasing the tipping diffusivity also. The situation in \cite{Patriarca.Leppaenen2004Modelinglanguagecompetition} corresponds to the case where the inter-community diffusivity is very low. The absence of perfect bilingualism as a stable solution is contrary to the results of \cite{Kosmidis.Halley.ea2005Languageevolutionand} in which almost all the simulations ended up with bilingualism. 

The present paper is not mainly dedicated to the language problem. The language problem is treated to show how the two-community model could be applied to other problems. It can be modified to incorporate the details that might be necessary for the specific field of application. For example, to realize the concept of a prestigious language \cite{Abrams.Strogatz2003Modellingdynamicsof}, an asymmetric community model will be necessary. Although an extremely asymmetric case were treated with a concept of tipping news \cite{Shin2010Tippingnewsin}, general cases could be treated with asymmetric inter-community diffusivities. A merit of the IAS model as applied to language competition problems is that it suggests the average size and content of vocabulary of the individual agents, which is never found in existing models. In its application to language problems, IAS shows possible stable outcomes in a single phase diagram. Refining the model remains a topic of future studies.

\section{Conclusion}
In this paper, we applied the IAS model to a two community problem which was previously studied in \cite{Lambiotte.Ausloos.ea2007Majoritymodelnetwork} with a different model of opinion formation. The main purpose of the paper was to identify the condition under which the two communities can reach consensus or maintain opposite opinions or different languages. The condition was found analytically for a system composed of homogeneous agents. While the study in \cite{Lambiotte.Ausloos.ea2007Majoritymodelnetwork} suggests that there is a tipping ratio of inter-community to intra-community links between the individuals above which the two societies cannot remain in opposite opinions, the present study shows that the tipping ratio also depends on the diffusivity, meaning the strength of interaction, between the individuals. The most important message of the present study is that the tipping ratio increases with the increase in the intra-community diffusivity. This implies that the increase in interaction between individuals caused by modern communication tools can serve to stabilize the coexistence mode rather than enhancing the convergence of different societies.  

\ack
J. K. Shin was supported by the Yeungnam University research grants in 2009.
J. Lorenz has received funding from the European Community's Seventh Framework Programme (FP7/2007-2013) under grant agreement no.~231323 (CyberEmotions project). We thank Frank Schweitzer (ETH Zurich) and two anonymous referees for valuable comments.

\appendix
\section{Information Accumulation System Model}\label{ap:2}

The version of the IAS model presented in Eq.~(\ref{eq:1}) is a simplified version of the general equation for the information of agent $i$ at time step $t$ as
\begin{equation}
 y_{i}^{t+1}=\tau_i y_{i}^t+(x_{i}^t+\sum_{j\in \Gamma_i}\lambda_i\omega_j y_{j}^t) (1-|y_{i}^t|) \label{eq:ap:1}
\end{equation}
Information of agent $i$ at time step $t+1$ is made up from three different contributions. The inheritance term $\tau_iy_{i}^t$  where the factor $\tau_i\leq 1$ is the individual \emph{inheritance rate} of agent $i$, representing personal memory. Part of the yesterday's information is carried to today. The counterpart of the inheritance rate $\tau_i$ is the \emph{volatility} $\Delta_i = 1-\tau_i$. It is set homogeneously equal to $\Delta$. The quantity  $x_{i}^t$ denotes the news and represents information from an external source, it is usually thought to be stochastic. In the present study, we only treat a small news problem in which we assume $x_{i}^t = 0$. The term $\sum_{j\in \Gamma_i} \lambda_i\omega_j y_{j,i}$ is the diffusivity term, which reflects the interactions between the agents in the system, with term $\omega_j$ is being the \emph{diffusivity} of agent $j$ and $\lambda_i$ being the absorption rate of agent $i$. In the present study $\omega_j$ is uniformly equal to $\omega$ and $\lambda_i$ is omitted. (Note, that one can also see it the other way round.) The factor ${(1-|y_{i}^t|)}$ is the saturation factor as in the simplified version. 

In the context of information seen as emotion the inheritance term reflects the fact that the intensity of the emotion should naturally decay towards the ground state zero if it is not further triggered by certain stimuli. These stimuli can be delivered in two ways, either externally by the news term which then represents external stimuli or by the emotions of its neighbors by ``emotional contagion'' (see \cite{Parkinson.Fischer.ea2005Emotioninsocial}). For example a facebook user can become sad about the death of Micheal Jackson mainly because all its friends are sad about it. This information can be propagated easily (often automatically) to a large number of people in modern communication platforms.

The IAS can successfully model situations in which a belief about a certain information can be sustained for a long time in a community. Examples are the long lasted geocentric theory of the universe until it was replaced by the heliocentric theory or the wide variety of cultural systems in communities that are sustaining for a long time through social inheritance. In terms of emotions the IAS can model self-enforcing collective emotions towards certain issues, which may have lead to long term behavioral norms, e.g. how snuffling is perceived in different cultural contexts, or what emotions colors of certain styles of clothing trigger.

\section{Nonsymmetric unstable fixed points in PN/NP-mode and full list of fixed points}\label{ap:1}
\newcommand{\totalomega}{\Omega_0+\Omega_X}
\newcommand{\omegadiff}{\Omega_0-\Omega_X}
\newcommand{\totalminusvolatility}{\totalomega-\Delta}
\newcommand{\insqrt}{\left(\frac{(\omegadiff)^2}{\totalomega} - \Delta\right) (\totalminusvolatility)}

When solving Eq.~(\ref{eq:3}) for fixed points of the PN/NP-mode, a quadruple of unstable fixed points may exist. 
Consider 
\begin{equation}
 Y_\mathrm{PN1} = \frac{(\omegadiff)(\totalminusvolatility)-(\totalomega)\sqrt{C}}{(\totalomega)(\totalminusvolatility)+(\totalomega)\sqrt{C}} \label{eq:A}
\end{equation}
and 
\begin{equation}
 Y_\mathrm{PN2} = \frac{(\totalminusvolatility) + \sqrt{C}}{2\Omega_0}\label{eq:B}
\end{equation}
with abbreviation
\[
C = \insqrt,
\]
then one can check that under the conditions \mbox{$\totalomega>\Delta$}, \mbox{$\Omega_0>\Omega_X$}, and $C>0$ it holds $Y_\mathrm{PN1},Y_\mathrm{PN2}>0$. If these conditions hold, then the fixed points $(A,-B)$ and $(B,-A)$ exist for the PN-mode and $(-Y_\mathrm{PN1},Y_\mathrm{PN2})$ and $(-Y_\mathrm{PN2},Y_\mathrm{PN1})$ for the NP-mode. These fixed points are also shown in the basins of attraction in Fig.~\ref{fig:2} (two basins of attraction on the right hand side).
One can also check that $C=0$ solved for $\Omega_X$ leads to Eq.~(\ref{eq:OmegaXC})  for the critical $\Omega^T$. Thus, these unstable fixed points exist when the skew-symmetric fixed points in PN/NP-mode are stable.

Table~\ref{tab:App} gives a full list of all fixed points and their conditions for existence, attractivity, and stability. This table accompanies Fig.~\ref{fig:2} to give a complete dynamical overview. The three horizontal lines between the fixed points in the table correspond to the three lines in the phase diagram in Fig.~\ref{fig:2} which are arranged clockwise around the point $(\Delta,0)$ (solid, dotted, solid) and to the three conditions in the table (i), (ii), and (iii). At every line a pitchfork bifurcation happens. First, the stable trivial solution undergoes a supercritical bifurcation into two new stable fixed points and gets unstable. For the second line let us focus on the anti-diagonal of the $(y_1,y_2)$-plane. On this line the trivial solution is still stable but after the second transition it undergoes a second supercritical pitchfork bifurcation into two new fixed points which are stable on the anti-diagonal (but unstable on the full plane) while the trivial solution turns unstable also on the anti-diagonal. Finally, at the third line the globally unstable fixed points on the anti-diagonal undergo subcritical bifurcations where they turn stable and each bifurcates into two new unstable fixed points. 

\begin{table}[h]
\centering
 \begin{tabular}{ccccc}
\textbf{mode}  & \textbf{position}                & \textbf{exists} & \textbf{unstable \& attractive}          & \textbf{stable \& attractive} \\ \hline\hline
00    & (0,0)                             & always & \parbox[t]{43mm}{\centering (i) \& $\neg$(ii) \\ \scriptsize{attractive on $y_1=-y_2$}}                  & $\neg$(i)        \\ \hline
PP    & $(Y_\mathrm{PP},Y_\mathrm{PP})$   & (i)    &  --                   & (i)                 \\
NN    & $(-Y_\mathrm{PP},-Y_\mathrm{PP})$ & (i)    &  --                   & (i)                 \\ \hline
PN    & $(Y_\mathrm{PN},-Y_\mathrm{PN})$  & (ii)   &  \parbox[t]{43mm}{\centering (ii), $\neg$(iii)\\ \scriptsize{attractive on $y_1=-y_2,y_1>0$}}    & (iii)          \\
NP    & $(-Y_\mathrm{PN},Y_\mathrm{PN})$  & (ii)   &  \parbox[t]{43mm}{\centering (ii), $\neg$(iii)\\ \scriptsize{attractive on $y_1=-y_2,y_2>0$}}    & (iii)          \\ \hline
PN    & $(Y_\mathrm{PN1},-Y_\mathrm{PN2})$                          & (iii)  &  --                   & --                  \\
PN    & $(Y_\mathrm{PN2},-Y_\mathrm{PN1})$                          & (iii)  &  --                   & --                  \\
NP    & $(-Y_\mathrm{PN1},Y_\mathrm{PN2})$                          & (iii)  &  --                   & --                  \\
NP    & $(-Y_\mathrm{PN2},Y_\mathrm{PN1})$                          & (iii)  &  --                   & --                  \\ \hline
\end{tabular} 
\vspace{3mm}

\fbox{\parbox[c]{0.39\textwidth}{
\textbf{Definitions}
\begin{compactitem}[]
\item $Y_\mathrm{PP} = 1-\frac{\Delta}{\Omega_0+\Omega_X}$
\item $Y_\mathrm{PN} = 1-\frac{\Delta}{\Omega_0-\Omega_X}$
\item For $Y_\mathrm{PN1},Y_\mathrm{PN2}$ see (\ref{eq:A}), (\ref{eq:B})
\end{compactitem}}}
\fbox{\parbox[c]{0.5\textwidth}{
\textbf{Conditions}
\begin{compactenum}[(i)]
\item $\Omega_0 + \Omega_X > \Delta$
\item $\Omega_0 - \Omega_X > \Delta$
\item $\insqrt > 0$
\end{compactenum}
For $\Omega_0,\Omega_X,\Delta>0$ it holds: (iii)$\Rightarrow$(ii)$\Rightarrow$(i) }}
\caption{Complete list of 9 possible fixed points and conditions for their existence, attractivity, and stability.}\label{tab:App}
\end{table}

\section*{References}
\bibliographystyle{unsrt}
\bibliography{refs}

\end{document}